\documentclass[twoside]{article}
\usepackage{fleqn,espcrc2}

\usepackage{graphicx}
\usepackage{psfig}
\usepackage{epsf}

\newcommand{\vrh}{\hat{\vr}}

\newcommand{\al}{\alpha}

\newcommand{\gm}{\gamma}

\newcommand{\ep}{\epsilon}

\newcommand{\kp}{\kappa}
\newcommand{\lm}{\lambda}
\newcommand{\rh}{\rho}

\newcommand{\ph}{\phi}
\newcommand{\vr}{\varphi}

\newcommand{\Sg}{\Sigma}

\newcommand{\half}{\frac{1}{2}}

\newcommand{\eela}[1]{\label{#1}\end{equation}}
\newcommand{\eeala}[1]{\label{#1}\end{eqnarray}}
\newcommand{\be}{\begin{equation}}
\newcommand{\ee}{\end{equation}}
\newcommand{\bea}{\begin{eqnarray}}
\newcommand{\eea}{\end{eqnarray}}

\newcommand{\ah}{\hat{a}}

\newcommand{\ahd}{\hat{a}^{\dagger}}

\newcommand{\AmS}{{\protect\the\textfont2
  A\kern-.1667em\lower.5ex\hbox{M}\kern-.125emS}}

\title{Damping and the Hartree Ensemble 
Approximation}

\author{Mischa Sall\'e\address{Institute
for Theoretical Physics, University of Amsterdam\\
Valckenierstraat 65, 1018 XE Amsterdam, the Netherlands}, 
Jan Smit\addressmark\thanks{Presented by J.~Smit}
and Jeroen C.~Vink\addressmark
}
       
\begin{document}

\begin{abstract}
We study a Hartree ensemble approximation for real-time
dynamics in the toy model of 1+1 dimensional scalar
field theory. Damping behavior seen in numerical simulations
is compared with analytical predictions based on perturbation
theory in the original (non-Hartree-approximated) model.
\vspace{1pc}
\end{abstract}

\maketitle

\section{Introduction}

Dynamics of quantum fields in real time is a lot more complicated than
statics in imaginary time and one has to make approximations
before giving the problem to the computer \cite{Boe}. As an improvement on 
classical dynamics we are studying the Hartree approximation.
The Hartree equations of motion for the $\vr^4$ model in 1+1 dimensions
are given by
\bea
\ddot \vr_x &=&\triangle\vr_x - ( \mu^2 + \lm \vr_x^2 + 3\lm C_x)\, \vr_x,
\nonumber\\
\ddot f_x^{\al} &=&\triangle f_x^{\al} -(\mu^2 + 3\lm \vr_x^2 + 3\lm C_x)\, 
f_x^{\al},
\nonumber\\
C_x &=& \sum_{\al} [(1 + 2n_0^{\al})|f_x^{\al}|^2],
\nonumber
\eea
where
$\vr_x=\langle\vrh_x\rangle$ is the mean field and the $f_x^{\al}$ 
are a complete set of mode functions used for the 
parametrisation of the operator field $\vrh_x$
in the Hartree approximation,
\[
\vrh_x = \vr_x + \sum_{\al} (\ah_{\al} f_x^{\al} + \ahd_{\al} f_x^{\al *}).
\]
The initial conditions are specified by a suitable choice for
$\vr_x$, $\dot\vr_x$
and the mode functions, as well as 
the initial occupation numbers
$n_0^{\al}=\langle\ahd_{\al}\ah_{\al}\rangle$.

\begin{figure}[htb]
\centerline{\psfig{figure=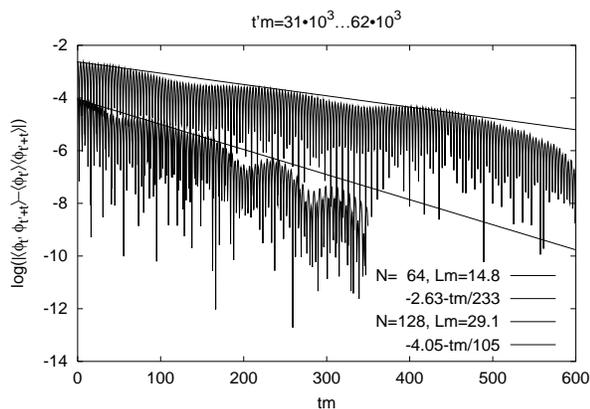,width=8 cm}}
\vspace{-0.5 cm}
\caption{Numerically computed auto-correlation functions $\ln|F_{\rm mf}(t)|$
versus time $mt$, with $m$ the temperature dependent mass. 
The coupling is weak, $\lm/m^2 = 0.11$ and the temperature $T/m\approx 1.4$
for the smaller volume (with significant deviations from the Bose-Einstein
distribution) and $\approx 1.6$ for the larger volume (reasonably BE).
}
\label{f1}
\vspace{-0.5 cm}
\end{figure}
Hartree-like approximations have been widely used, but 
when the mean field is homogeneous  
they do not lead to proper thermalization.
This may be ascribed to lack of
sufficient scattering in the interaction term involving a spatially
constant $C_x$. We try to improve on this \cite{Vi} by writing a homogeneous
initial density matrix as a superposition of gaussian pure states.
The Hartree approximation is then applied to 
realizations of this ensemble, which are
typically inhomogeneous. This allows for scattering of the modes 
(`particles') via the fluctuating mean field and redistribution
over various momenta.
To test for thermalization we start out of equilibrium with 
`the modes in their vacuum' ($n_0^{\al} = 0$) 
and compute equal-time- and auto-correlation functions.
For the resulting particle distributions, see \cite{Vi}.
Here we concentrate on damping phenomena.

We found funny modulations in auto-correlation functions. 
Fig.~\ref{f1} shows such modulations on top of a roughly exponential decay. 
The correlation function 
corresponds to the zero momentum mode of the mean field,
\[
F_{\rm mf}(t) =\int dx\;\overline{\vr(x,t)\vr(0,0)}_{\mbox{ }\rm conn},
\]
where the over-bar denotes a time average, 
taken after waiting a long time 
for the system to be in approximate equilibrium. 
As discussed in \cite{Vi}, this equilibrium
is approximately thermal. The function 
$F_{\rm mf}(t)$ is analogous to the symmetric correlation function 
of the quantum field theory:
\be
F(t) = \int dx\;\langle\half\{\vrh(x,t),\,\vrh(0,0)\}\rangle_{\mbox{ }\rm conn}.
\ee
Does
$F(t)$
also have the modulations? 

\section{Calculation of $F(t)$}
\begin{figure}[htb]
\centerline{\psfig{figure=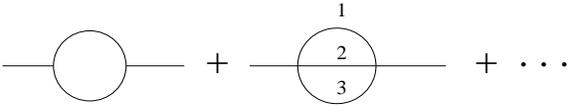,width=7.6cm}}
\caption{Diagrams leading to thermal damping.}
\label{f2}
\end{figure}

The function
$F(t)$
can be expressed in terms of the (zero momentum) spectral function
$\rh(p^0)$,
\be
F(t) = \int_{-\infty}^{\infty} \frac{dp^0}{2\pi}\, 
e^{-ip^0 t}\,
\left(\frac{1}{e^{p^0/T}-1}+\half\right)\, \rh(p^0)
\label{Fdef}
\ee
which is determined by the retarded selfenergy
$\Sg(p^0)$,
\be
\rh(p^0)=
\frac{-2{\rm Im}\,\Sg(p^0)}{[m^2 - p_0^2 + {\rm
Re}\,\Sg(p^0)]^2 + [{\rm Im}\,\Sg(p^0)]^2}.
\label{rhdef}
\ee
%
The selfenergy can be calculated in perturbation theory. 
The relevant diagrams are shown in Fig.~\ref{f2}.
The one-loop diagram is present only in the `phase of broken symmetry'.
It leads to damping for frequencies
$p_0^2> 4 m^2$, 
which are irrelevant for the quasiparticle damping at
$p_0^2 = m^2$. 
(There is really only a symmetric phase in 1+1 dimensions, 
but this is due to symmetry restoration by nonperturbative effects
which will not obliterate the damping mechanism described by the diagrams.)
So we concentrate on the two-loop diagram. 
This is given by the sum of two terms,
$\Sg_1 + \Sg_2$.
The first has an imaginary part corresponding to $1\leftrightarrow 3$
processes, requiring $p_0^2 > 9m^2$,
so it does not contribute to plasmon damping.  The second is given by
\bea
\Sg_2 &=&-\frac{9\lm^2}{16\pi^2}\int 
\frac{dp_2\, dp_3}{E_1E_2E_3}\,
\nonumber\\&&
\frac{
(1+n_1)n_2 n_3-n_1 (1+n_2) (1+n_3)}{p^0+i\ep +E_1 - E_2 - E_3} 
\nonumber\\&&
+ \left[(p^0 +i\ep)\to-(p^0+i\ep)\right],
\label{Sg2}
\eea
where
$\lm$
is the coupling constant (introduced as
${\cal L}_1 = -\lm\vr^4/4$
), and
\bea
E_1 &=& \sqrt{m^2 + (p_2 + p_3)^2},\;\;
E_{2,3} = \sqrt{m^2 + p_{2,3}^2},
\nonumber\\
n_i &=& [\exp(E_i/T)-1]^{-1},
\;\;\;\; i=1,2,3.
\nonumber
\eea
Its imaginary part corresponds to
$2\leftrightarrow 2$
processes, which contribute to plasmon damping (the regions near
$p_0=\pm m$
).

\begin{figure}
\centerline{\psfig{figure=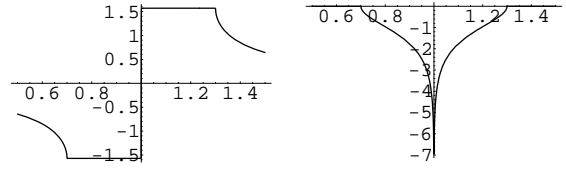,width=7.8cm}}
\caption{Real (left) and imaginary (right) part of
$-i\ln\left(i/z + \sqrt{1-1/z^2}\right) = \mbox{arcsin}(1/z)$,
for
$z=(p^0 + i\ep-m)/(0.13 m)$ and
$\ep/m = 10^{-4}$.
}
\label{f3}
\end{figure}
Now the formula for the thermal plasmon damping
rate (at zero momentum) in terms of the retarded selfenergy, 
\be
\gm = -{\rm Im}\, \Sg(m)/2m, 
\label{gmdef1}
\ee
leads to a
{\em logarithmically divergent}
answer. This is a collinear divergence which is absent 
in more than one space dimension.
Inspection shows that the singular part of
$\Sg_2$
is given by the non-relativistic region
of the integral in (\ref{Sg2}). 
Using polar coordinates 
$p_2 = p\cos\ph$, $p_3 = p\sin\ph$ 
this non-relativistic 
($p<\kp\ll m$) contribution is 
proportional to 
\[
\int_0^{\kp}p\, dp\, \int_0^{2 \pi}\frac{d\ph}{2\pi}\,
\frac{1}{p^0 + i\ep - m + (p^2/2m)\sin 2\ph}
\]
\be
\nonumber\\
=  m \left[-i
\ln\left(\frac{i}{z}+\sqrt{1-\frac{1}{z^2}}\right)\right],
\label{Sgnr}
\ee
\[
z = \frac{p^0 + i\ep -m}{\kp^2/2m},
\]
where $p^0\approx m$.
This function is plotted in Fig.\ \ref{f3}.
\begin{figure}[t]
\centerline{\psfig{figure=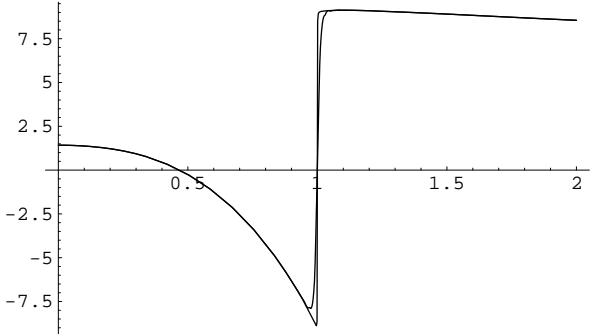,width=7.8cm}}
\caption{Plot of
${\rm Re}\, \Sg(p^0)/(9\lm^2/16\pi^2)$
obtained by linear extrapolation
$\ep= 0.02$, 0.01
to zero, together with a matching to the logarithmic singularity
($T=m=1$).
}
\label{f4}
\end{figure}
\begin{figure}[t]
\centerline{\psfig{figure=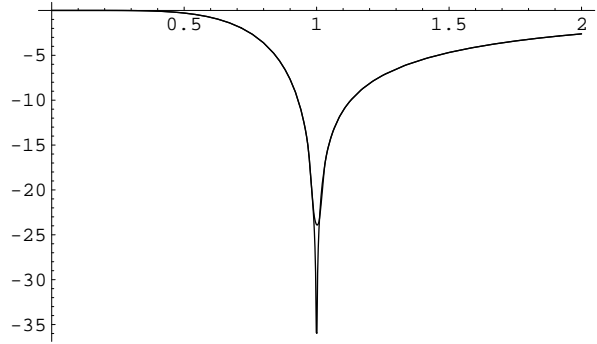,width=7.8cm}}
\caption{As in Fig. 4 for the imaginary part.}
\label{f5}
\end{figure}

A natural way out of the difficulty posed by the divergent
$\Sg(m)$
may be to continue the selfenergy
analytically into the lower half of its second Riemann sheet,
$p^0\to m-i\gm$,
and replace (\ref{gmdef1}) by the improved definition
\be
m^2 - (m-i\gm)^2 + \Sg(m-i\gm) = 0.
\label{gmdef2}
\ee
For weak coupling
$\lm/m^2\ll 1$
we then get 
the equation
\be
\frac{\gm}{m} = \frac{9\lm^2}{16\pi m^4}\, 
\frac{e^{m/T}}{\left(e^{m/T}-1\right)^2}\,
\left[\ln\frac{m}{\gm}+c(T)\right],
\label{gmeq}
\ee
where the constant
$c$ 
has to be determined by matching a numerical evaluation of
$\Sg$
to the form (\ref{Sgnr}) for
$p^0 \approx m$.

We evaluated
$\Sg_2$
in (\ref{Sg2}) for
$T=m$
by numerical integration with
$\ep/m = 0.02$, $0.01$
and linear extrapolation
$\ep\to 0$.
The result is shown in Figs.\ 4 and 5,
together with a matching to the logarithmic
singularity, giving
$c\approx -0.51$. 
For example, 
Eq.\ (\ref{gmeq}) now gives
$\gm/m = 0.061$, 
for
$\lm/m^2 = 0.4$. 

To see how well this
$\gm$ 
describes the decay of the correlator
$F(t)$
we evaluated this function
directly from (\ref{Fdef}) and (\ref{rhdef}).
The divergence in
${\rm Im}\,\Sg(p^0)$
at
$p^0 = m$
leads to a 
{\em zero}
in the spectral function
$\rh(p^0)$.
So is there a peak at all in
$\rh(p^0)$?
Fig.\ 6 shows what happens: 
the `usual' peak has separated into two twins!

Fig.\ 7 shows the resulting
$F(t)$. 
The effect of the double peak is indeed an oscillating modulation on top
of the roughly exponential decay. The decay corresponding to 
$\exp(-\gm t)$,
with
$\gm$
given by (\ref{gmeq}),
is also indicated in the plot: it
does {\em not} do a good job in describing the average decay beyond the first
interference minimum. 
The `Twin Peaks' phenomenon implies
that the usual definition of damping rate (\ref{gmdef2})
is unreliable in 1+1 dimensions.
\begin{figure}[t]
\centerline{\psfig{figure=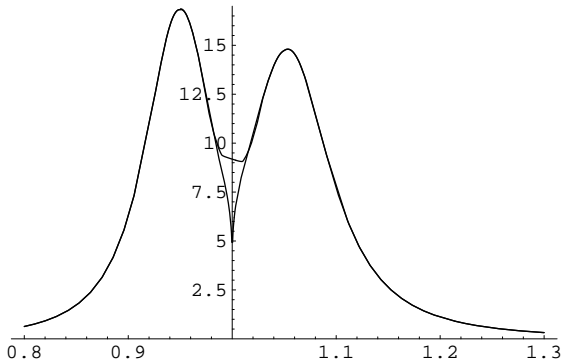,width=7.8cm}}
\caption{The spectral function $\rh(p^0)$ near $p^0 = m =1$
corresponding to the selfenergy shown in Figs.\ 4, 5
($T=m$, $\lm=0.4\, m^2$).
}
\label{f6}
\end{figure}
\begin{figure}
\centerline{\psfig{figure=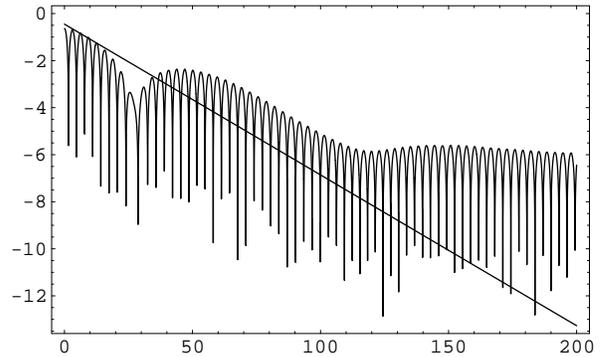,width=7.8cm}}
\caption{Plot of $\ln|F(t)|$ versus $mt$ for $T=m$, $\lm= 0.4\, m^2$.
}
\label{f7}
\end{figure}
Fig.~8 shows the result of a calculation of $F(t)$ with 
parameters taken from the numerical simulation in Fig.~1 
with the larger volume. In this case $\ep$ was kept finite,
$\ep/m=0.005$, which may be more realistic since one expects anyway the 
infinity in ${\rm Im}\, \Sg$ to be smeared out by damping effects in the
propagators in Fig.~\ref{f2}.
Figs.\ 1 and 8 are reasonably similar, but how to compare the 
the average slopes is somewhat ambiguous.
\begin{figure}[b]
\centerline{\psfig{figure=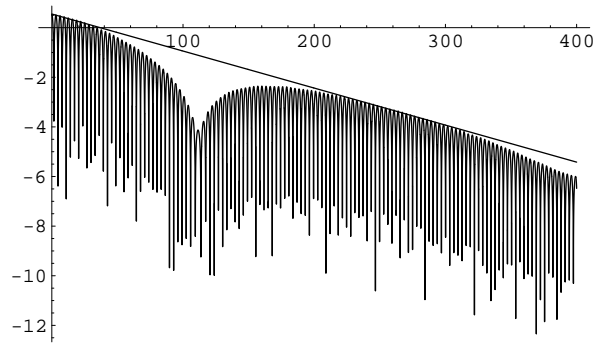,width=7.8cm}}
\caption{$\ln|F(t)|$ versus $mt$ with  
$\lm/m^2 = 0.11$, $T/m=1.63$, 
corresponding to Fig.~1.
The line is given by $\exp(0.55-t/67)$.
}
\label{f8}
\end{figure}

\section{Outlook}
Summarizing, we are encouraged by the similarities in the
qualitative features of the numerical and analytical auto-correlation
functions. Quantitatively, the damping times are also of the same order
of magnitude (105 vs 67 $m^{-1}$), but we did not really accurately 
compute the relevant auto-correlation
function yet in our simulations, which
will require a lot more numerical effort.

\section*{Acknowledgements} We thank Gert Aarts for useful conversations. 
This work is supported by FOM/NWO.

\end{document}